\documentclass[pdflatex,sn-mathphys-num]{sn-jnl}

\usepackage{graphicx}%
\usepackage{multirow}%
\usepackage{amsmath,amssymb,amsfonts}%
\usepackage{amsthm}%
\usepackage{mathrsfs}%
\usepackage[title]{appendix}%
\usepackage{xcolor}%
\usepackage{textcomp}%
\usepackage{manyfoot}%
\usepackage{booktabs}%
\usepackage{algorithm}%
\usepackage{algorithmicx}%
\usepackage{algpseudocode}%
\usepackage{listings}%

\usepackage{tabularx} 

\usepackage[utf8]{inputenc}
\usepackage[table]{xcolor}
\definecolor{matchGray}{gray}{0.9}

\definecolor{tableHeader}{RGB}{52, 73, 94}
\definecolor{tableRowOne}{RGB}{245, 247, 249}
\definecolor{tableRowTwo}{RGB}{255, 255, 255}

\usepackage{natbib}
\usepackage{graphicx}
\usepackage{caption}
\usepackage{xurl} 
\usepackage{hyperref}
\usepackage{xcolor}
\usepackage{todonotes}

\theoremstyle{thmstyleone}%
\theoremstyle{thmstyletwo}%

\theoremstyle{thmstylethree}%

\begin{document}

\title[Article Title]{T-TExTS (\textbf{T}eaching \textbf{T}ext \textbf{Ex}pansion for \textbf{T}eacher \textbf{S}caffolding): Enhancing Text Selection in High School Literature through Knowledge Graph-Based Recommendation}


\author[1]{\fnm{Nirmal} \sur{Gelal}}\email{ngelal@ksu.edu}

\author[2]{\fnm{Chloe} \sur{Snow}}\email{chlo24@ksu.edu}

\author[2]{\fnm{Ambyr} \sur{Rios}}\email{ambyrrios@ksu.edu}

\author[1]{\fnm{Kathleen M.} \sur{Jagodnik}}\email{kmjagodnik@ksu.edu}

\author*[1]{\fnm{Hande Küçük} \sur{McGinty}}\email{hande@ksu.edu}

\affil[1]{\orgdiv{Department of Computer Science}}

\affil[2]{\orgdiv{Department of Curriculum and Instruction}}

\affil[]{\orgname{Kansas State University}, \orgaddress{\city{Manhattan}, \postcode{66502}, \state{KS}, \country{USA}}}


\abstract{
High school English Literature teachers often encounter barriers to assembling diverse, thematically aligned text sets due to limited planning time and pedagogical resources. To address this need, we present \mbox{T-TExTS} (Teaching Text Expansion for Teacher Scaffolding), a knowledge graph (KG)-based recommendation system that suggests literature texts based on pedagogical merit rather than surface-level metadata. We construct a domain-specific ontology using the Knowledge Acquisition and Representation Methodology (KNARM), instantiate it as a knowledge graph with separate Terminological Box (TBox) and Assertional Box (ABox) components, and evaluate four graph embedding strategies (DeepWalk, biased random walk, hybrid embedding, and Node2Vec) across three dataset configurations (98, 196, and 351 texts) and two relation-weighting schemes. The experimental results reveal that traversal-level expert weighting alone does not outperform algorithmic structural tuning: Node2Vec achieves the highest Area Under the Curve (AUC) at every dataset size (0.9642--0.9750) and the strongest ranking metrics (Hits@K, MRR, nDCG) at larger scales. Combining structural and pedagogical signals through embedding concatenation, however, preserves both interpretability and competitive ranking quality, with the hybrid model maintaining a high AUC across all scales (0.9122--0.9350) and remaining within a few percentage points of Node2Vec on every ranking metric. These findings highlight the value of ontology-driven knowledge graph embeddings for educational recommendation systems and demonstrate that \mbox{T-TExTS} can meaningfully ease the burden of English Literature text selection for secondary educators, supporting more informed and inclusive curricular decisions. The source code for \mbox{T-TExTS} is available at \url{https://github.com/koncordantlab/TTExTS}.
}




\maketitle

\section{Introduction}
\label{sec:introduction}

Scholars argue that ``transformative pedagogy" in the secondary classroom requires a multiliteracy approach \cite{jacobs2012proverbial}. Text sets are one form of multiliteracy: groups of textual works in any format, thematically aligned, that are taught together to develop a specific concept or deepen students' understanding of a topic \cite{opatz2022building}. They support instruction by building background knowledge, improving text accessibility, and scaffolding comprehension \cite{elish2014scaffolding,lupo2018building}. Beyond academic benefits, text sets also expose students to literary and informational works that are culturally relevant and produce engaging reading experiences \cite{elish2014scaffolding}. They thus give secondary teachers a means to bring texts of varied genre, author, length, complexity, and modality into their curricula \cite{lupo2020rethinking}.
 
Despite these benefits, many educators do not realize the full value of text sets because they lack the planning time, materials, or pedagogical training needed to assemble and implement them in their curriculum \cite{lupo2018building,lupo2020rethinking}. This study addresses the need for an automated scaffolding system that helps educators to select literature texts that are diverse in genre, theme, subtheme, and author, yet remain pedagogically coherent. We present Teaching Text Expansion for Teacher Scaffolding (\mbox{T-TExTS}), a recommendation system that suggests high school English Literature texts based on pedagogical merit, genre, and thematic relevance using a domain-specific knowledge graph (KG).

\begin{figure}[htbp]
\centering
\captionsetup{justification=raggedright, singlelinecheck=false}
\includegraphics[width=0.7\columnwidth]{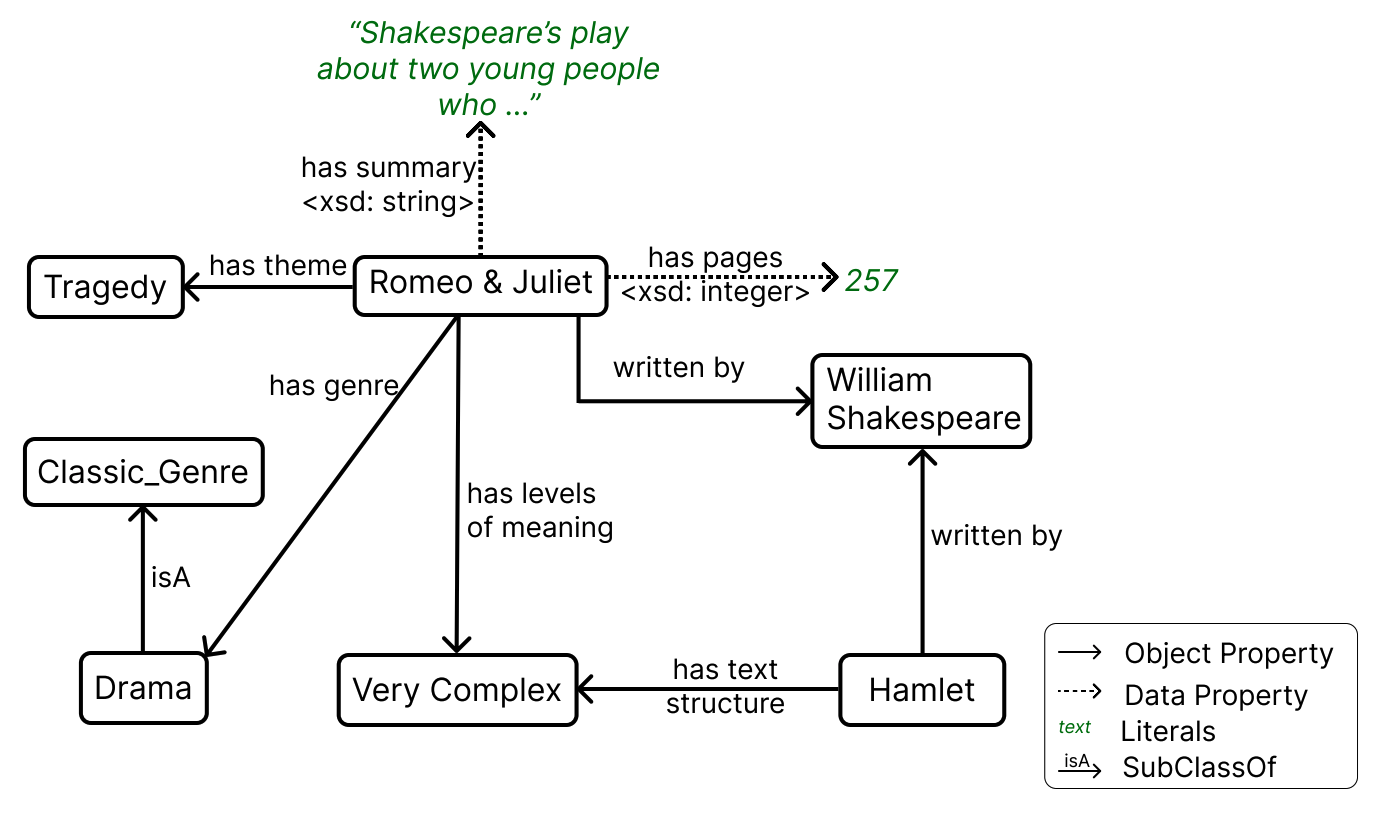} 
\caption{\textit{Example of a knowledge graph representing English Literature concepts and their relationships.}}
\label{fig1}
\end{figure}

To support this pedagogical scaffolding, we use knowledge graphs grounded in a domain-specific ontology. Unlike traditional relational databases, KGs formally model the semantic relationships among entities, such as texts, themes, and authors, enabling the system to represent the interconnected nature of literary concepts. Figure \ref{fig1} illustrates the basic structure of a knowledge graph. By applying graph embedding techniques, specifically random-walk-based methods such as DeepWalk \cite{deepwalkperrozi} and Node2Vec \cite{node2vec}, we project these symbolic relationships into a continuous vector space. This transformation enables \mbox{T-TExTS} to quantify the ``pedagogical proximity" between texts based on graph structure rather than keyword overlap, allowing the system to identify recommendations that are thematically and contextually aligned with a teacher's anchor text.

Traditional recommendation approaches such as Collaborative Filtering (CF) and Content-Based Filtering (CBF) have shown effectiveness \cite{ricci2010introduction}, but each has notable limitations. CF often suffers from the cold-start problem, where sparse data on new entities degrades recommendation quality \cite{thorat2015survey,yang2018novel}. Standard CBF systems, in turn, struggle to extend beyond a user's existing interests and rarely introduce new or related topics. KG-based recommenders address both limitations \cite{guo2020survey} by encoding semantic relationships between entities, supporting more personalized, context-aware recommendations even when interaction data are sparse, and surfacing related content beyond a user's existing preferences. While KGs have been successfully applied in domains such as \mbox{e-commerce} and movie recommendation \cite{guo2020survey,rosati2016rdf}, \mbox{T-TExTS} is novel in adopting a pedagogy-first ontology that explicitly prioritizes educational scaffolding over surface-level content similarity or popularity.

The contributions of this paper are threefold. First, we propose a pedagogy-grounded ontology for high school English Literature, constructed using the Knowledge Acquisition and Representation Methodology (KNARM) \cite{knarm}. Second, we develop \mbox{T-TExTS}, a KG-based recommendation system that uses pedagogical attributes to suggest thematically aligned text sets. Third, we conduct a comparative evaluation of four graph embedding strategies on \mbox{T-TExTS} (DeepWalk, biased random walk, hybrid, and Node2Vec) across three dataset configurations and two relation-weighting schemes (default and genre-emphasized), including a sensitivity analysis of how weight choice affects ranking and link-prediction performance.

The remainder of this paper is organized as follows. Section~\ref{sec:related_work} reviews related work on recommendation systems based on knowledge graphs and ontologies. Section~\ref{sec:methodology} presents our methodology, including data collection in subsection~\ref{sec:data_collection}, ontology creation and knowledge graph generation in subsection~\ref{sec:ontology_creation}, and the recommendation model in subsection~\ref{sec:recommendation_model}. Section~\ref{sec:empirical_experiments_and_performance_evaluation} reports the experiments and performance evaluation of the system. Section~\ref{sec:discussion_and_limitation} presents the discussion in subsection~\ref{sec:discussion} and limitations in subsection~\ref{sec:limitations}. Section~\ref{sec:conclusion_and_future_work} closes the paper with concluding remarks in subsection~\ref{sec:conclusion} and future directions in subsection~\ref{sec:future_work}.

\section{Related Work}
\label{sec:related_work}

The selection of classroom instructional texts by English Language Arts teachers has been described as significant and consequential for students' skill development and interest in reading \cite{robertson2020keeping, tan2021reader}. In particular, teachers' selection of high-quality, diverse texts that are relevant to and representative of the learners in their care has been linked to a range of social and academic student benefits \cite{graff2022contemporary, yokota2017striving}. Despite these potential benefits and the growing number of resources for diverse text selection, the texts that teachers actually select have remained largely stagnant over the past century \cite{rios2024kansas, watkins2015navigating}. Teachers continue to choose texts with which they are familiar, those available within their schools, and those within the secondary English Literature canon \cite{rios2024kansas}, all of which yield a group of texts that is more traditional, less diverse, and often a mismatch with students' identities and interests \cite{ervin2022critically, glaws2021conversations, ivey2001just}. Teachers, therefore, require additional tools and resources to support the creation and integration of text sets in the secondary English Language Arts curriculum.

Representation learning over ontologies \cite{smaili2018onto2vec,smaili2019opa2vec} and knowledge graphs \cite{chen2020graph,yi2022graph,khoshraftar2024survey} has been studied extensively. Graph representation learning supports a range of downstream tasks, including link prediction \cite{kumar2020link}, node classification \cite{bhagat2011node}, relation prediction \cite{rossi2021knowledge}, and recommendation \cite{guo2020survey,yang2022knowledge}. \citet{agibetov2018fast} and \citet{alshahrani2017neuro} applied feature learning to biological knowledge graphs and reported improvements in link-prediction performance. These results show that knowledge graphs perform well across different domains, capturing both implicit and explicit domain knowledge for the tasks listed above.

\citet{rosati2016rdf} used general knowledge graphs such as DBpedia and Wikidata for content-based recommendation of movies and books. Their work applied a language-modeling approach for unsupervised feature extraction from sequences of words, adapted to Resource Description Framework (RDF) graphs. The performance of this language-modeling approach on domain-specific knowledge graphs, however, was not tested. In addition, \citet{rosati2016rdf} relied only on uniform random walks for sequence generation, whereas our approach evaluates four embedding strategies on a domain-specific KG: a uniform random walk (DeepWalk), an expert-guided biased random walk, a parameterized random walk (Node2Vec), and a hybrid that concatenates the uniform and biased embeddings.

\citet{cochez2017biased} proposed 12 strategies for biased random walks based on uniform, edge-centric, node-centric object frequency, and node-centric PageRank approaches, evaluating them across five datasets from different domains. We do not adopt those specific strategies; instead, our method is guided by the importance of predicates for our particular task, as identified by domain experts. Their work nevertheless provided useful direction and identified areas for improvement.

Node2Vec \cite{node2vec}, a widely used random-walk-based embedding algorithm, introduced a biased walk procedure controlled by two parameters: the return parameter $p$ and the in-out parameter $q$. These parameters allow the random walk to balance exploration strategies by interpolating between Breadth-First Search (BFS) and Depth-First Search (DFS). BFS captures structural equivalence by focusing on immediate local structure, while DFS explores broader neighborhoods and aids the discovery of homophilous communities at the cost of higher variance. Given its effectiveness and widespread adoption, we include Node2Vec among the four embedding strategies evaluated in this study, alongside DeepWalk, a domain-expert-guided biased random walk, and a hybrid embedding that combines uniform and biased representations.

Another approach to learning features from a graph relies on graph neural networks (GNNs), which are based on the concept of \textit{message passing}. In a GNN, message passing refers to the process by which information, represented as vectors, is exchanged between nodes along edges. \citet{de2024personalized} applied a GNN to personalized audiobook recommendations on the Spotify platform. The work of \citet{kipf2016semi} and \citet{velivckovic2017graph} reported improved performance on prediction and classification tasks over traditional graph feature learning algorithms, introducing architectures such as Graph Convolutional Networks (GCNs) and Graph Attention Networks (GATs). UI-KCGN \cite{gu2025does} uses a GCN to show the value of integrating user-side knowledge graphs to enhance personalization based on user attributes. Our work differs by focusing only on a high-fidelity, expert-curated ontology and random-walk embeddings to enforce pedagogical coherence, rather than user-similarity signals. Although GNNs such as GCNs and GATs have shown strong performance on various graph-based tasks, our current study uses random-walk-based approaches because high-complexity neural architectures are prone to overfitting when applied to expert-curated knowledge graphs whose scale is bounded by manual annotation effort \cite{dwivedi2023benchmarking}.

While KGs have been increasingly adopted in educational recommender systems, existing research has focused mainly on concept sequencing and learning-path optimization in Science, Technology, Engineering, and Mathematics (STEM) domains or Massive Open Online Courses (MOOCs) \cite{chen2021learning, chen2022recommending, cen2025egrec}. A distinct gap remains in the application of KGs to literacy education and teacher-facing tools. Unlike concept-based domains where recommendations follow a linear prerequisite logic (e.g., Algebra I $\rightarrow$ Algebra II), literature recommendations require capturing non-linear, semantic connections between texts, such as shared themes, social justice attributes, and genre complexity. To our knowledge, \mbox{T-TExTS} is the first system to apply KG-based recommendation specifically to the pedagogical scaffolding of text sets for high school English Literature.

\section{Methodology}
\label{sec:methodology}

\subsection{Data Collection}
\label{sec:data_collection}
The initial set of English Literature texts in this work was drawn from findings of a study on text selection practices conducted across the state of Kansas \cite{rios2024kansas}. Additional texts were added through two approaches: (1) consultation with educators and (2) analysis of higher-education course syllabi. To supplement the dataset through educator input, we solicited guidance from middle and high school teachers in local Kansas districts who reported using MyPerspectives\footnote{\href{https://www.savvas.com/solutions/literacy/core-programs/myperspectives-english-language-arts}{https://www.savvas.com/solutions/literacy/core-programs/myperspectives-english-language-arts}} and StudySync,\footnote{\href{https://www.studysync.com/}{https://www.studysync.com/}} two widely adopted, standards-aligned English Language Arts (ELA) curriculum programs used in K-12 schools across the state. These educators provided information on the literary texts currently used in their classrooms. In parallel, we examined syllabi for higher-education courses in Kansas State University's Department of English, specifically ENGL 545: Literature for Adolescents and ENGL 580: World Literature, developed by department faculty. These syllabi were used to identify additional literature frequently assigned in secondary and postsecondary educational settings.

For each selected text, quantitative analyses were performed using established text-complexity tools, including the Lexile Analyzer,\footnote{\url{https://hub.lexile.com/text-analyzer/}} the Flesch-Kincaid Grade Level Calculator,\footnote{\url{https://goodcalculators.com/flesch-kincaid-calculator/}} and the ATOS Analyzer.\footnote{\url{https://www.renaissance.com/resources/atos-analyzer/atos-analyzer-tool/}} These standardized metrics evaluate the complexity of a text based on linguistic features such as sentence length, word frequency, and syntactic structure, yielding an estimate of its difficulty. In addition to these quantitative measures, qualitative evaluations were conducted using rubrics from the Achieve the Core initiative,\footnote{\label{SCASS}\href{https://www.scribd.com/doc/259926327/scass-text-complexity-qualitative-measures-lit-rubric-2-8}{SCASS Text Complexity Qualitative Measures Rubric}} aligned to the \textit{Research Supporting Key Elements of the Standards, Common Core State Standards for English Language Arts and Literacy in History/Social Studies and Science and Technical Subjects} \cite{nysedAppendixA}. We applied this procedure across three dataset configurations of increasing scale, comprising 98, 196, and 351 English Literature texts together with their associated qualitative, quantitative, and instructional merit descriptors. These three configurations are used in the scaling study reported in Section~\ref{sec:empirical_experiments_and_performance_evaluation}. To support validity, results were corroborated within the text-evaluation research team.

Text summaries were derived from descriptions available through the Library of Congress database.\footnote{\url{https://catalog.loc.gov}} To determine genre, theme, and sub-theme classifications, we used keywords from the same database and identified emergent elements within each subcategory. We labeled the category as \textit{genre}, the major takeaways from the text as \textit{theme}, and the descriptive, emotive topics related to the text as \textit{sub-theme}. We also constructed a taxonomy-like hierarchical tree for genre, theme, and sub-theme, which was later preserved as separate modules within the ontology.

\subsection{Ontology Creation}
\label{sec:ontology_creation}

Ontologies offer a formal mechanism to define concepts (classes), properties (relationships), and individuals within a specific domain, and serve as a structured means of transforming qualitative domain knowledge into a machine-interpretable representation. Ontologies employ formal languages, often based on Description Logics (DLs), to precisely capture meanings and logical associations among real-world entities. A standard convention in DL is to separate the schema-level Terminological Box (TBox), which defines class hierarchies and properties, from the instance-level Assertional Box (ABox), which records facts about specific individuals. In this study, the ontology and the resulting knowledge graph were derived from curated educational texts and pedagogical criteria used in text selection, with close involvement of domain experts throughout the process. The TBox encodes the schema for high-level concepts such as \texttt{Text}, \texttt{Author}, \texttt{Genre}, \texttt{Theme}, and \texttt{TextComplexity}, while the ABox holds per-text assertions linking each text and author as individuals through relations such as \texttt{has\_author}, \texttt{has\_genre}, and \texttt{has\_theme}.

We used a data-agnostic methodology, known as \textbf{KN}owledge \textbf{A}cquisition and \textbf{R}epresentation \textbf{M}ethodology (KNARM) \cite{knarm}, to generate our ontology. Figure \ref{fig2} illustrates the nine steps and overall flow of KNARM, which guide the identification of relevant literary concepts, pedagogical attributes, and their relationships from the text selection process. Following Agile Principles, feedback loops are employed prior to the finalization of ontologies. The circular flow depicted in the figure signifies that the ontology-building process constitutes a continuous effort, enabling ontology engineers to iteratively incorporate additional concepts and knowledge until the captured knowledge is validated and approved by domain experts.\footnote{Throughout this work, the term \textit{Domain Experts} denotes the current secondary English Language Arts teachers and teacher educators who are researchers specializing in text selection and curriculum development for secondary school settings.} The nine steps of the KNARM methodology, along with the corresponding actions performed in each step, are briefly outlined below:

\begin{figure}[htbp]
\centering
\captionsetup{justification=raggedright, singlelinecheck=false}
\includegraphics[width=0.6\columnwidth]{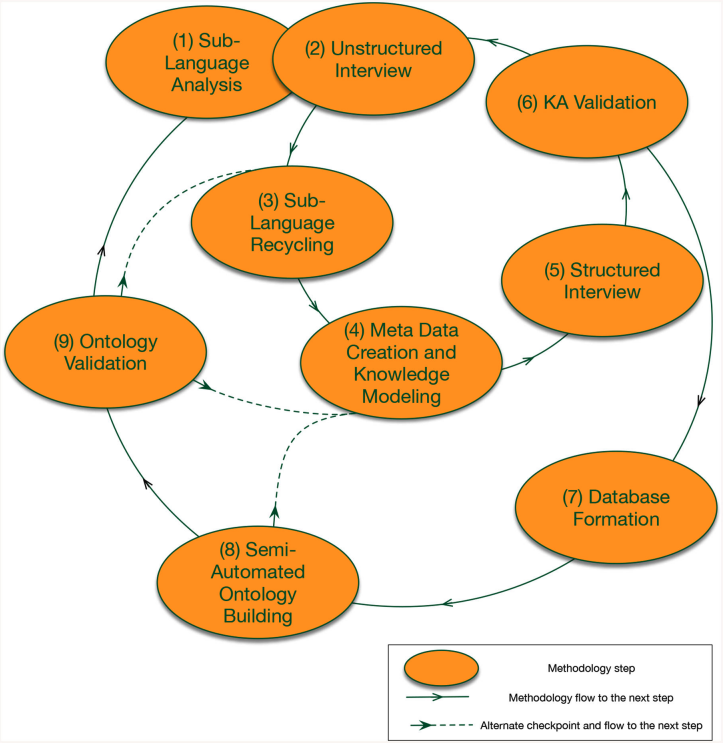} 
\caption{\textit{Steps of KNowledge Acquisition and Representation Methodology (KNARM) \cite{knarm}, where KA refers to Knowledge Acquisition.}}
\label{fig2}
\end{figure}

\subsubsection{Sub-Language Analysis}
In this first step, we analyzed the key concepts and relationships associated with text characteristics, pedagogical value, and diversity in terms of genre, theme, and subtheme within the English Literature curriculum domain. A variety of document corpora, including the \citet{nysedAppendixA} and the SCASS Text Complexity Qualitative Measures Rubric\footref{SCASS} were analyzed to understand the qualitative measures and used to distinguish the complexity of texts. Recurring terms and patterns related to text complexities, pedagogical merits,\footnote{Pedagogical merits are evaluated through a combination of qualitative measures—including levels of meaning, text structure, language conventionality and clarity, and knowledge demands—and quantitative metrics, including Lexile Level, Flesch-Kincaid Grade Level, ATOS Reading Level, and Reader Maturity Level.} genre, theme, and sub-theme were identified, which constituted the aim of this step. Following the identification of concepts and their relationships, validation was conducted with domain experts to verify and refine them.

\subsubsection{Unstructured Interviews}
We conducted several unstructured interviews with literacy teachers, curriculum developers, and English Language Arts subject matter experts who select texts for high schools. We employed open-ended questions exploring how they currently select English Literature texts, what defines a ``diverse" \footnote{``Diverse" texts here refer to texts that differ in terms of genre, theme, subtheme, and authors, yet have similar pedagogical merits.} text, how they assess similarity in complexity or pedagogical approach, and their expectations for this recommendation system.

\subsubsection{Sub-Language Recycling}
Manual curation of data was performed, as there are, to the best of our findings, neither existing datasets nor ontologies available that contain pedagogical merits and literary elements for English Literature. The detailed process of using existing relevant documents in order to create vocabulary for an ontology is described in section \ref{sec:data_collection}.

\subsubsection{Metadata Creation and Knowledge Modeling}
\begin{figure}[htbp]
\centering
\captionsetup{justification=centering}
\includegraphics[width=\columnwidth]{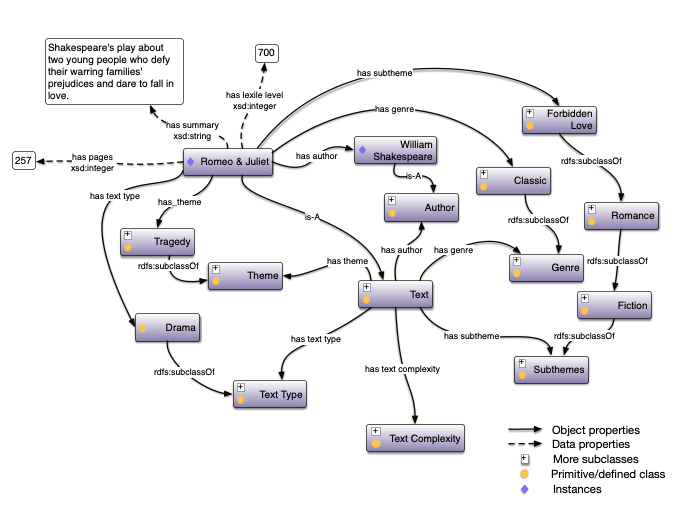} 
\caption{\textit{High-level schema of the \mbox{T-TExTS} ontology}}
\label{fig3}
\end{figure}

We identified the essential concepts (e.g., Book, Author, Genre, Theme) and their relationships (e.g., hasAuthor, hasGenre, hasTheme) to construct a metadata framework describing the domain of the data to be modeled. Metadata, in this context, refers to elements that define or describe what constitutes a text. Knowledge modeling plays a vital role in this step, in conjunction with metadata creation. Figure \ref{fig3} presents a high-level representation of a schema of the ontology that was modeled in this step.

\subsubsection{Structured Interviews}
The knowledge obtained in the previous steps was used to conduct interviews with domain experts involved in data creation. These interviews were structured to ensure that no important aspects were overlooked during the modeling process. We added a few rule-based axioms in this step, such as: if the year of a text is earlier than 1945, then the triple (Text, hasEra, Traditional) is generated, as it describes texts written before the end of World War II \cite{rybakova2016connecting}; else, the triple (Text, hasEra, Contemporary) is generated. Additionally, if any of the qualitative measures contain ``slightly complex" as a field and the Lexile level is less than 925, then the triple (Text, hasTextComplexity, Slightly Complex) is included in the graph. Similarly, for Lexile levels greater than or equal to 925 and less than 1185, with any of the qualitative measures being ``moderately complex," a triple (Text, hasTextComplexity, Moderately\_Complex) is added. Texts with Lexile level ranges 1185 to 1335 were designated as ``very complex", and 1335 to 1440 as ``exceedingly complex" texts. Likewise, through continuous interaction with the domain experts, we collected the knowledge for each text and associated concepts.

\subsubsection{Knowledge Acquisition Validation}
This phase serves as the initial feedback loop in the process. Outputs from previous steps, including the modeled knowledge, were presented to domain experts. The primary aim of this step was to identify any gaps in the knowledge that may have been missed or misinterpreted.

\subsubsection{Database Formation}
While our datasets are small enough to be efficiently managed using flat files, we generated a GraphDB repository containing the TBox schema and the ABox assertions for each dataset configuration to ensure rapid additions and revisions to the knowledge graph in the future. Graph databases, such as GraphDB and Neo4j, offer numerous advantages, including handling large-scale data, structural compatibility with the schema, scalability, and their inherent graph-based architecture.

\subsubsection{Semi-Automated Ontology Building}
We used the RDFLib\footnote{\url{https://github.com/RDFLib/rdflib}} library to create and modify the knowledge graphs. This step is referred to as semi-automated because, while the schema development and data population are automated using these tools, domain experts remain actively engaged throughout the process to ensure accuracy and relevance.

\subsubsection{Ontology Validation}
Diverse approaches are available to validate ontologies. \citet{raad2015survey} propose four different approaches to ontology evaluation: gold standard-based, corpus-based, task-based, and criteria-based evaluation. Since there exist no ontologies in high-school-level English Literature texts containing pedagogical merits for gold standard-based evaluation, nor any document corpus that comprehensively captures all the concepts and relationships defined and incorporated in our ontology and knowledge graphs for corpus-based evaluation, we adopted a task-based ontology evaluation mechanism, which is also better suited than criteria-based evaluation to the purpose-built nature of our ontology. Our task-based ontology validation approach involves measuring the performance of an ontology for a particular task, regardless of the ontology's structural characteristics. This approach focuses on addressing the domain-expert requests identified during the interview phases of ontology development.

\begin{figure*}[htbp]
\centering
\captionsetup{justification=centering}
\includegraphics[width=\columnwidth]{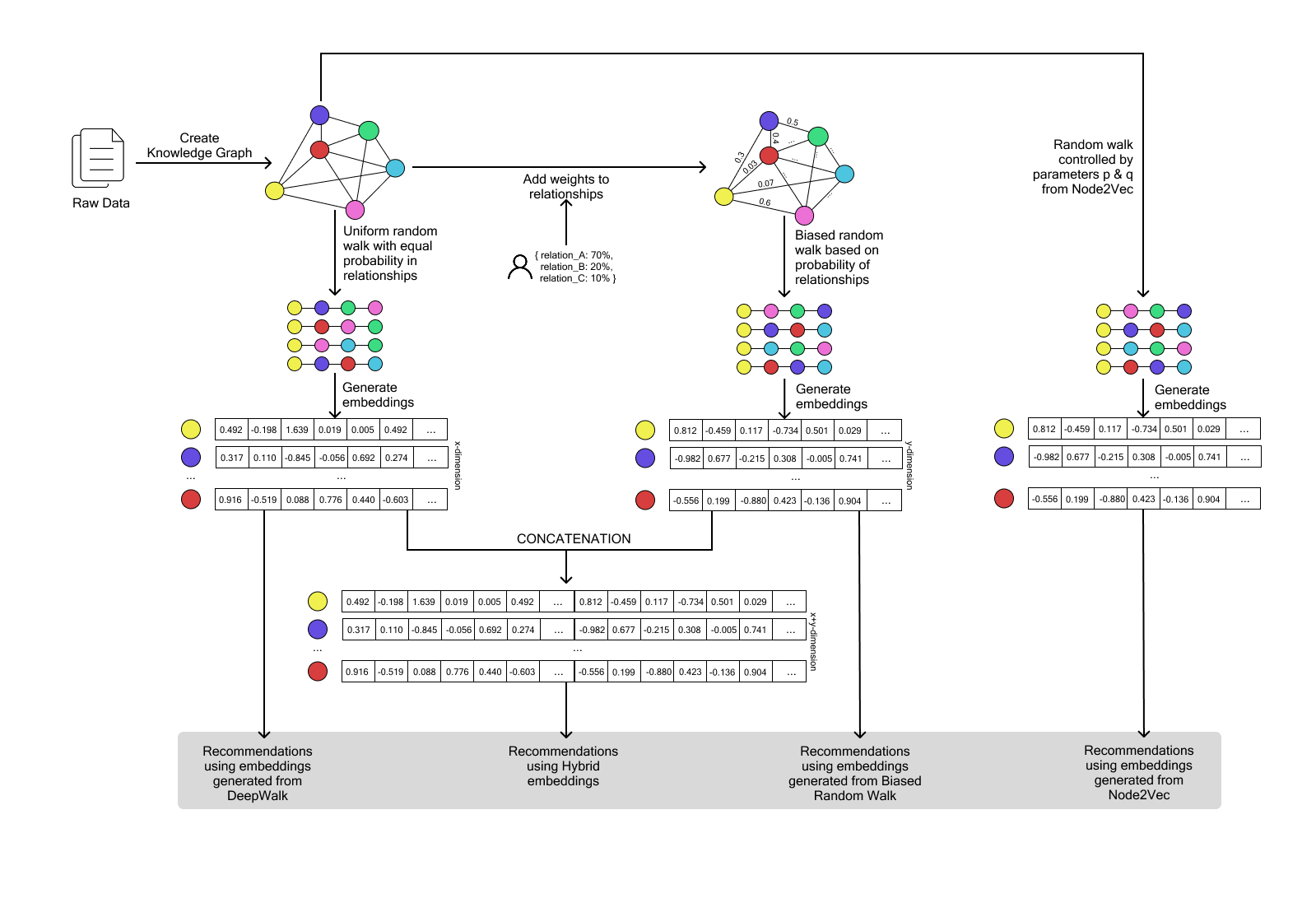} 
\caption{\textit{Workflow of the \mbox{T-TExTS} English Literature Text Recommender System}}

\captionsetup{justification=justified, font=small}
    \caption*{\footnotesize{Raw educational text data are transformed into a domain-specific knowledge graph using the KNARM methodology \cite{knarm}. Random-walk-based embedding methods are then applied, including uniform random walks (DeepWalk), domain-expert-guided biased random walks, and parameterized random walks (Node2Vec \cite{node2vec}). These walks generate vector representations of graph entities, which are used either directly or combined via embedding concatenation to form hybrid representations. Finally, similarity computation over the embeddings produces pedagogically relevant text recommendations, and the resulting recommendations from different embedding strategies are compared.}}

\label{fig4}
\end{figure*}

\subsection{Recommendation Model}
\label{sec:recommendation_model}

Machine learning (ML) applications and statistical analyses utilize numerical data, making the conversion of text- and graph-based data into vector representations a common approach. \citet{chen2021owl2vec} advocate two paradigms of representing entities in a vector space while preserving their semantics. The first approach involves computing embeddings in an end-to-end manner, iteratively adjusting the vectors using an optimization algorithm to minimize the overall loss across all the triples. Losses are typically calculated by scoring the truth or falsity of each triple (positive and negative samples). Algorithms such as DistMult \cite{distmult} and TransE \cite{transE} are based on this technique.

Another paradigm involves explicitly exploring neighborhoods of entities and relations in the graph prior to learning embeddings using a word embedding model. DeepWalk \cite{deepwalkperrozi} adopts this strategy by employing a random walk technique to explore neighborhoods and generate sequences of nodes (and relations in KGs), treating these sequences as ``sentences''. For each concept and relation within the sentence, a latent feature or embedding is learned using the skip-gram method described in Word2Vec \cite{mikolov2013efficient}. 

In this work, we adopt the latter paradigm due to its flexibility in incorporating traversal bias. The ontology $\mathcal{O}$ is serialized into a graph $\mathcal{G}$ of RDF triples, on which random-walk-based embedding methods can be applied directly. Let $\mathcal{E}$ = \{$e_1,\ldots,e_{N_e}$\} represent the set of entities (concepts and individuals), and let $\mathcal{P}$ = \{$p_1,\ldots,p_{N_p}$\} in $\mathcal{G}$ denote the set of predicates (relations) connecting two entities. A walk $\mathcal{W}_{e_i}$ is defined as an ordered sequence of entities $\{e_i, e_j, e_k, \ldots\}$, generated by traversing the graph according to a specified random-walk strategy. For simplicity, walks are represented as a sequence of entities; relations influence traversal probabilities, but are not explicitly encoded in the walk sequences. Depending on the embedding method, the transition probabilities governing the walk may be uniform, relation-weighted, or parameterized. In the uniform setting, each neighboring entity is selected with equal probability. In the weighted variant, transition probabilities are adjusted based on the relative importance of relations, as defined by domain experts. In parameterized random walks, such as those employed by Node2Vec, exploration behavior is controlled through return and in-out parameters, allowing interpolation between local and global neighborhood traversal. As the focus of this study is on relational structure among entities, data properties, i.e., relationships between an entity and a literal value such as strings or numerical attributes, were excluded from the graph prior to walk generation.

Using these walk strategies, we generate sequences of entities that are treated as sentences and embedded using the skip-gram model \cite{mikolov2013efficient} implemented in Word2Vec. Given a target entity $w_t$, the objective of the skip-gram model is to maximize the probability of observing surrounding context entities within a fixed window size $c$:
\begin{gather*}
    \text{maximize} \prod_{-c\leq j\leq c,\, j\neq 0} P(w_{t+j}\mid w_t)
\end{gather*}
This process yields dense vector representations for entities such that entities appearing in similar structural contexts are embedded closer together in the vector space.

Embedding quality benefits from explicit contrast between observed and unobserved triples \cite{negsampling}, since the model learns to distinguish true relations from false ones. To enable this contrast, we generated negative samples using a standard negative-sampling technique. During negative sample generation, either the subject (i.e., head) or the object (i.e., tail) of a triple is replaced by another entity, ensuring that the resulting triple does not exist in the graph. For example, from the true triple \textit{\textless Romeo\&Juliet, written\_by, William\_Shakespeare\textgreater}, negative triples are generated by substituting either the head \textit{Romeo\&Juliet} or the tail \textit{William\_Shakespeare} with other entities. The resulting negative triples may include \textit{\textless Romeo\&Juliet, written\_by, George\_Orwell\textgreater} or \textit{\textless 1984, written\_by, William\_Shakespeare\textgreater}, which are false. To prevent the accidental inclusion of true triples as negative examples, we implemented a filtering constraint: every candidate negative triple $(h',r,t)$ or $(h,r,t')$ was cross-referenced against the set of all positive triples in the original knowledge graph. If a generated triple was found to exist in the knowledge graph, it was immediately discarded, and a new replacement was sampled until a unique, non-existent triple was obtained. A selected subset of negative samples was manually reviewed to check the relevance to the scope. This negative sample generation technique has a few limitations. For instance, for a triple \textit{\textless Romeo\&Juliet, isA, Text\textgreater}, changing a head to another Text that is not in the knowledge graph would classify this triple to be a negative sample. In an open-world assumption, what is not known to be true may still be true. Since knowledge graphs follow the open world assumption, we cannot imply that the resulting negative triple is false. However, we adopt the Local Closed World Assumption (LCWA) during training to provide the model with a necessary contrastive signal, effectively forcing the embedding space to concentrate probability mass on observed triples while penalizing unobserved ones. Using this negative sample generation technique, the size of the graph was increased, and the dataset with added negative samples was used as the training dataset.

To assess embedding quality prior to recommendation, we first tuned the model hyperparameters using the link prediction task. Hyperparameters of the models, including walk length, number of walks per node, embedding dimensionality, and context window size, were tuned using an automatic hyperparameter optimization framework, Optuna\footnote{\href{https://github.com/optuna/optuna}{https://github.com/optuna/optuna}}. For Node2Vec, the return parameter $p$ and in-out parameter $q$ are also tuned to balance local and global exploration. The optimization process was guided by link prediction performance, measured using the Area Under the Curve (AUC) of the Receiver Operating Characteristic (ROC). AUC assesses the model's ability to rank true triples higher than negative samples and provides an effective proxy for embedding quality prior to recommendation.

In addition to evaluating individual embedding strategies, we introduce a hybrid representation to examine whether complementary structural signals can be combined. Specifically, embeddings generated using uniform random walks (DeepWalk) and domain-expert-guided biased random walks are concatenated to form a hybrid embedding. This design is motivated by the observation that uniform walks capture broad structural context, while biased walks emphasize pedagogically salient relations. Node2Vec embeddings are evaluated independently as a strong baseline; however, they are not included in the hybrid representation to avoid redundancy, as Node2Vec generalizes uniform random walks through parameterized exploration.

Figure~\ref{fig4} illustrates the complete workflow of the proposed system, including knowledge graph construction, random-walk-based embedding generation, embedding fusion, and recommendation. The biased random walk strategy is further detailed in Algorithm~1. In this approach, for an entity $e_i$, the generated walk will be \{$e_i, e_j,...$\} if the predicate between $e_i$ and $e_j$ has a greater weight than any of the predicates connecting $e_i$ to its other neighboring nodes. The next node $u_{next}$ is selected randomly based on the normalized probability, which modifies the original uniform random walk and aims to generate more focused walks. One of the major advantages of the biased random walk approach is that the walks can be adjusted based on concepts and relationships of interest. For instance, if the objective is to recommend a text primarily based on genre, the weight of the edges representing genre can be maximized. This produces sequences with greater representation of genres and texts as entities, allowing the embedding model to capture more relevant information.

\begin{algorithm}
\caption{Biased random walk with weighted transition}
\begin{algorithmic}[0]
\Require Graph $G$, start node $u_0$, walk length $L$, relation weights $\mathcal{R}$
\Ensure Random walk path $[u_0, u_1, \ldots, u_k]$
\State Initialize walk $W \gets [u_0]$
\For{$i = 1$ to $L - 1$}
    \State $u \gets$ last node in $W$
    \State $\mathcal{N} \gets$ neighbors of $u$ in $G$
    \If{$\mathcal{N} = \emptyset$}
        \State \textbf{break}
    \EndIf
    \State $u_{\text{next}} \gets \textsc{WeightedChoice}(\mathcal{N}, \mathcal{R})$
    \State Append $u_{\text{next}}$ to $W$
\EndFor
\State \Return $W$

\vspace{5pt}\hrule\vspace{5pt}
\Statex \textbf{function} \textsc{WeightedChoice}($\mathcal{N}, \mathcal{R}$)
\Statex \hspace{1em} $\mathcal{R}_{\text{default}} \gets \min(\mathcal{R}.\text{values}())$
\Statex \hspace{1em} $p \gets [\mathcal{R}.\text{get}(relation(u,v), \mathcal{R}_{\text{default}}) \text{ for } v \in \mathcal{N}]$
\Statex \hspace{1em} Normalize $p \gets p / \sum(p)$
\Statex \hspace{1em} \Return sample from $\mathcal{N}$ with weights $p$

\end{algorithmic}
\end{algorithm}

The embeddings of structurally related entities are positioned close to one another in the high-dimensional vector space, as the embedding algorithms are designed to preserve the topology of the graph. The proximity between two embeddings can be quantified using similarity measures such as Cosine Similarity, Euclidean Distance, Manhattan Distance, and Jaccard Similarity. In this work, the Cosine Similarity technique was employed, which measures the cosine of the angle between two non-zero vectors and emphasizes the direction of the vectors rather than their magnitude. The similarity range is $[-1,1]$, where 1 indicates that the two vectors are identical, 0 denotes orthogonality (no similarity or correlation), and -1 signifies that the vectors are diametrically opposite. For any two vectors $A$ and $B$, cosine similarity is given by:
\begin{gather*}
    cos(\theta) = \frac{A\cdot B}{\|A\|\|B\|}
\end{gather*}

For a given text $e_k$ and its embedding $\Vec{e_k}$, cosine similarity was applied to all other texts to extract the $top\_n$ recommendations. 

\section{Empirical Experiments and Performance Evaluation}
\label{sec:empirical_experiments_and_performance_evaluation}

A domain-specific knowledge graph was constructed to model the relationships and entities within the defined domain of high-school-level English Literature texts. To evaluate the robustness of the proposed approach as the graph grows in entity count and relational density, we conducted a scaling study across three dataset configurations containing 98, 196, and 351 texts. Each configuration was curated in close collaboration with domain experts following the KNARM-guided protocol described in Section~\ref{sec:methodology}. The key structural statistics for all three knowledge graphs are summarized in Table~\ref{tab:kg-stats}.

\begin{table}[htbp]
\centering
\caption{Structural statistics of the \mbox{T-TExTS} knowledge graphs at three dataset sizes. Entity counts reflect the ABox-centric construction described in Section~\ref{sec:ontology_creation}, in which texts and authors are represented as individuals rather than subclasses.}
\label{tab:kg-stats}
\begin{tabular}{lrrr}
\toprule
\textbf{Feature} & \textbf{98 texts} & \textbf{196 texts} & \textbf{351 texts} \\
\midrule
Total triples (with literals)     & 3,972 & 6,448 & 10,331 \\
Triples without literals          & 3,302 & 5,241 & 8,248  \\
Unique entities                   & 868   & 1,053 & 1,312  \\
Classes                           & 600   & 600   & 600    \\
Object properties                 & 13    & 13    & 13     \\
Data properties                   & 8     & 8     & 8      \\
Avg.\ relationships per entity    & 3.97  & 5.20  & 6.53   \\
\bottomrule
\end{tabular}
\end{table}

The constant counts of class, object-property, and data-property across the three dataset sizes reflect that the TBox is shared across all configurations; only the ABox grows as the dataset expands. The increasing average relationships per entity indicate that the larger datasets are not just more numerous but also more richly interconnected, providing a stronger test of the embedding models' ability to capture relational structure.

\subsection{Experimental Setup}
\label{sec:experimental-setup}
For each dataset size, the knowledge graph was split into training, validation, and test sets using an 80/10/10 ratio. The split sizes for the three configurations are reported in Table~\ref{tab:splits}. Negative samples were generated using the procedure described in Section~\ref{sec:recommendation_model}, with head-or-tail corruption followed by a filtering constraint that cross-references each candidate negative triple against the full set of positive triples to prevent accidental inclusion of true triples as negative examples.

\begin{table}[htbp]
\centering
\caption{Edge split sizes for training, validation, and testing across the three \mbox{T-TExTS} dataset configurations.}
\label{tab:splits}
\begin{tabular}{lrrrr}
\toprule
\textbf{Dataset} & \textbf{Total edges} & \textbf{Training} & \textbf{Validation} & \textbf{Test} \\
\midrule
98  & 3,302 & 2,641 & 330 & 331 \\
196 & 5,241 & 4,192 & 524 & 525 \\
351 & 8,248 & 6,598 & 824 & 826 \\
\bottomrule
\end{tabular}
\end{table}

Hyperparameters for each embedding model (DeepWalk, biased random walk, and Node2Vec) were tuned independently on each dataset size using the Optuna optimization framework\footnote{\url{https://github.com/optuna/optuna}}, with the validation split used to compute link-prediction AUC as the optimization objective. A total of 50 trials were conducted per model and dataset size. All random number generators, including Optuna's sampler, were seeded to ensure reproducibility across runs. The best hyperparameters for each configuration are reported in Table~\ref{tab:hyperparams}.

\begin{table}[htbp]
\centering
\caption{Best hyperparameters identified by Optuna for each embedding model in \mbox{T-TExTS} across dataset sizes. The hybrid model does not involve additional tuning; it is formed by concatenating embeddings from DeepWalk and a biased random walk. Biased random walk hyperparameters shown here correspond to the default weighting configuration; the genre-emphasized configuration used in the sensitivity analysis (Section~\ref{sec:sensitivity}) uses separately tuned hyperparameters.}
\label{tab:hyperparams}
\begin{tabular}{llrrrrrr}
\toprule
\textbf{Model} & \textbf{data\_size} & \textbf{walk\_len} & \textbf{num\_walks} & \textbf{dim} & \textbf{window} & \textbf{p} & \textbf{q} \\
\midrule
\multirow{3}{*}{DeepWalk}  & 98  & 80  & 40 & 512  & 30 & -- & -- \\
                           & 196 & 40  & 40 & 1024 & 25 & -- & -- \\
                           & 351 & 100 & 10 & 128  & 25 & -- & -- \\
\midrule
\multirow{3}{*}{Node2Vec}  & 98  & 80  & 20 & 512  & 20 & 1.00 & 4.0 \\
                           & 196 & 100 & 50 & 1024 & 30 & 2.00 & 4.0 \\
                           & 351 & 80  & 50 & 1024 & 30 & 0.25 & 4.0 \\
\midrule
\multirow{3}{*}{Biased random walk} & 98  & 100 & 50 & 512  & 30 & -- & -- \\
                           & 196 & 80  & 20 & 1024 & 30 & -- & -- \\
                           & 351 & 60  & 25 & 1024 & 25 & -- & -- \\
\bottomrule
\end{tabular}
\end{table}

A notable pattern emerges in the Node2Vec hyperparameters: the in-out parameter $q$ consistently converges to $4.0$ across all dataset sizes, indicating a strong preference for DFS-biased exploration that moves away from immediately-visited neighborhoods. The implications of this finding for knowledge graphs over literary corpora are discussed in Section~\ref{sec:discussion}.

\subsection{Evaluation Metrics}
Hits@K, Mean Reciprocal Rank (MRR), and normalized Discounted Cumulative Gain (nDCG@10) were used to evaluate the ranking performance of the recommendation system. Hits@K measures whether the correct item appears within the top-K predicted results; the final score represents the fraction of test cases in which the correct answer appears in the top-K:
\[
    \text{Hits@K} = \frac{\text{\# of correct items in top-$K$}}{\text{\# of test cases}}
\]
 
MRR measures the average quality of ranks by considering the reciprocal of the rank of the first correct answer. Its value ranges from 0 (no correct answers) to 1 (all correct items ranked at the top):
\[
    \text{MRR} = \text{Average}\left(\frac{1}{\text{First correct rank}}\right)
\]
 
Normalized Discounted Cumulative Gain (nDCG) measures the extent to which the ordering of the items returned by the system aligns with the ordering of items that an ideal system would return. It is defined as the ratio of the system's Discounted Cumulative Gain (DCG) at rank $K$ to the Ideal Discounted Cumulative Gain (IDCG) at the same rank, where IDCG is the DCG that would be obtained if the items were perfectly sorted by relevance:
\begin{align*}
    \text{DCG@K}   &= \sum_{i=1}^{K} \frac{\text{rel}_i}{\log_2(i+1)} \\
    \text{IDCG@K}  &= \sum_{i=1}^{K} \frac{\text{rel}_i^{\text{sorted}}}{\log_2(i+1)} \quad \text{(sorted descending)} \\
    \text{nDCG@K}  &= \frac{\text{DCG@K}}{\text{IDCG@K}}, \quad \text{ranges } [0, 1]
\end{align*}

Link prediction was assessed using the Area Under the Curve (AUC) score, computed as the area under the Receiver Operating Characteristic (ROC) curve. AUC ranges from 0 to 1, where a score of 1 indicates perfect performance and a score below 0.5 is worse than random. Because AUC measures the model's ability to rank positive cases higher than negative ones without being sensitive to absolute candidate-pool size, it serves as a scale-independent measure of embedding quality.

\subsection{Recommendation Performance Across Dataset Sizes}
\label{sec:results-main}
Table~\ref{tab:main-results} presents the performance of DeepWalk, biased random walk (with default pedagogy-balanced weights), the hybrid embedding approach, and Node2Vec on the task of recommending pedagogically aligned high school English Literature texts, evaluated across the three dataset sizes.

\begin{table}[htbp]
\centering
\caption{Performance of \mbox{T-TExTS} on the high school English Literature text recommendation task, comparing four embedding strategies (DeepWalk, biased random walk, hybrid, and Node2Vec) across three dataset sizes. Bolded values indicate the best performance within each row. Biased random walk results correspond to the default weighting configuration; the genre-emphasized sensitivity study is reported separately in Table~\ref{tab:sensitivity}.}
\label{tab:main-results}
\begin{tabular}{llrrrr}
\toprule
\textbf{Dataset Size} & \textbf{Metric} & \textbf{DeepWalk} & \textbf{Biased RW} & \textbf{Hybrid} & \textbf{Node2Vec} \\
\midrule
\multirow{7}{*}{98}
  & AUC       & 0.9653          & 0.6888 & 0.9175          & \textbf{0.9750} \\
  & Hits@1     & \textbf{0.5263} & 0.3684 & 0.4211          & 0.4211          \\
  & Hits@3     & \textbf{0.5789} & 0.5263 & 0.5263          & 0.5263          \\
  & Hits@5     & 0.5789          & 0.5789 & \textbf{0.6316} & \textbf{0.6316} \\
  & Hits@10    & 0.6842          & 0.6842 & \textbf{0.7368} & 0.6842          \\
  & MRR       & \textbf{0.5604} & 0.4630 & 0.5004          & 0.5026          \\
  & nDCG@10   & \textbf{0.5167} & 0.4762 & 0.5131          & 0.5150          \\
\midrule
\multirow{7}{*}{196}
  & AUC       & 0.9595          & 0.7107 & 0.9122          & \textbf{0.9642} \\
  & Hits@1     & 0.1842          & 0.2105 & \textbf{0.2368} & 0.2105          \\
  & Hits@3     & \textbf{0.4211} & 0.3421 & 0.3947          & \textbf{0.4211} \\
  & Hits@5     & 0.5000          & 0.4474 & 0.4211          & \textbf{0.5789} \\
  & Hits@10    & \textbf{0.6842} & 0.6316 & 0.6316          & 0.6579          \\
  & MRR       & 0.3372          & 0.3215 & 0.3548          & \textbf{0.3602} \\
  & nDCG@10   & 0.4034          & 0.3737 & 0.3922          & \textbf{0.4269} \\
\midrule
\multirow{7}{*}{351}
  & AUC       & 0.9611          & 0.7258 & 0.9246          & \textbf{0.9730} \\
  & Hits@1     & \textbf{0.1250} & 0.0938 & 0.0938          & \textbf{0.1250} \\
  & Hits@3     & 0.2969          & 0.2344 & 0.2656          & \textbf{0.3281} \\
  & Hits@5     & 0.3750          & 0.3281 & 0.3438          & \textbf{0.4062} \\
  & Hits@10    & 0.5156          & 0.4531 & 0.5156          & \textbf{0.6094} \\
  & MRR       & 0.2422          & 0.2013 & 0.2191          & \textbf{0.2503} \\
  & nDCG@10   & 0.2921          & 0.2399 & 0.2760          & \textbf{0.3262} \\
\bottomrule
\end{tabular}
\end{table}

Four patterns emerge across the three dataset sizes.

\paragraph{AUC is stable across scales.}
DeepWalk maintains an AUC between 0.9595 and 0.9653 across the three dataset sizes, and Node2Vec between 0.9642 and 0.9750. This stability indicates that the structural signal captured by uniform and parametrically guided random walks generalizes well as the knowledge graph grows. All four models operate on the same expert-curated ontology, so the pedagogical structure is encoded at the graph level for every method; what differs is whether traversal further amplifies specific relations through expert-assigned weights. The biased random walk does, while DeepWalk and Node2Vec do not. The biased random walk maintains a substantially lower AUC (0.6888 to 0.7466), reflecting the trade-off inherent in its constrained exploration: while the additional traversal-level weighting concentrates walks on relations that domain experts identified as most pedagogically informative, it sacrifices coverage of the graph's global topology, which AUC as a link-prediction metric rewards.

\paragraph{Ranking metrics decrease as the candidate pool expands.}
Hits@K, MRR, and nDCG@10 all decline as dataset size increases, which is expected. Retrieving the exact top-K matches becomes strictly harder as the number of candidate texts grows, and the relative ranking of correct items also degrades because the position of any single ground-truth match becomes less informative within a larger pool. For instance, DeepWalk Hits@1 drops from 0.5263 at 98 texts to 0.1250 at 351 texts, and DeepWalk nDCG@10 drops from 0.5167 to 0.2921 over the same range. This is not a deficiency of the models, but rather a property of these metrics: Hits@K, MRR, and nDCG@10 all measure retrieval quality over an increasingly large set, and their absolute values are therefore not directly comparable across dataset sizes. The stability of AUC, which is not sensitive to absolute candidate-pool size, provides a clearer signal of representation quality across scales. Practically, this means that a tool based on \mbox{T-TExTS} would present teachers with a ranked list at every corpus size, from which near-miss recommendations remain pedagogically valuable even when the single top-ranked item is not one of the expert-approved ground-truth matches.

\paragraph{No single embedding method dominates all metrics at all scales.}
DeepWalk achieves the best MRR (0.5604) and Hits@1 (0.5263) at the 98-text scale, reflecting the effectiveness of unbiased traversal on small, dense graphs. As the dataset grows, Node2Vec takes the lead on ranking metrics: at 196 and 351 texts, Node2Vec achieves the highest MRR (0.3602, 0.2503) and nDCG@10 (0.4269, 0.3262). The hybrid model, which concatenates DeepWalk and biased random walk embeddings, occupies a middle ground: it matches or surpasses DeepWalk on several Hits@K metrics (for example, Hits@10 = 0.7368 at 98 texts) while preserving a high AUC across all scales (0.9122 to 0.9246). This trade-off profile is discussed further in Section~\ref{sec:discussion}.

\paragraph{The biased random walk underperforms on AUC but remains competitive on specific ranking metrics.}
Although the biased random walk has the lowest AUC at every scale, it matches or exceeds Node2Vec on Hits@1 at 196 texts (0.2105) and achieves competitive Hits@5 and Hits@10 values at several scales. This pattern reflects a trade-off between structural coverage and ranking precision: traversal-level expert weighting reduces the breadth of graph exploration, but improves accuracy on queries for which the weighted relations are most relevant. The hybrid model addresses both objectives by combining the biased random walk with uniform exploration.

\subsection{Sensitivity Analysis: Effect of Genre-Weighted Traversal}
\label{sec:sensitivity}

To assess the sensitivity of the biased random walk to weight configuration, we evaluated two weight schemes. The \textbf{default} configuration assigns equal weight (3) to \texttt{has\_genre}, \texttt{has\_theme}, and \texttt{has\_subtheme} relations, with qualitative measures (such as \texttt{has\_levels\_of\_meaning} and \texttt{has\_text\_structure}) weighted at 2, and all other relations at 1. The \textbf{genre-emphasized} configuration retains this scheme, but increases the weight of \texttt{has\_genre} to 4, elevating it above theme and subtheme. This ablation is motivated by pedagogical design choices: while diversity across themes and subthemes is a curricular goal, genre is often the first organizing dimension that teachers use when assembling a text set. The question being examined here is whether privileging genre during graph traversal improves the resulting recommendations.

Table~\ref{tab:sensitivity} presents the ablation results. Because the hybrid model concatenates biased random walk embeddings with DeepWalk embeddings (which are weight-independent), we report both the biased random walk and hybrid model under each weight configuration.

\begin{table}[htbp]
\centering
\caption{Sensitivity of \mbox{T-TExTS} high school English Literature text recommendation performance to weight configuration. Default and genre-emphasized weight configurations are compared for the biased random walk and hybrid models. Bolded values indicate the better of the two weight configurations within each model-size pair.}
\label{tab:sensitivity}
\begin{tabular}{llrrrr}
\toprule
 & & \multicolumn{2}{c}{\textbf{Biased RW}} & \multicolumn{2}{c}{\textbf{Hybrid}} \\
\cmidrule(lr){3-4} \cmidrule(lr){5-6}
\textbf{Dataset Size} & \textbf{Metric} & \textbf{Default} & \textbf{Genre-emp.} & \textbf{Default} & \textbf{Genre-emp.} \\
\midrule
\multirow{7}{*}{98}
  & AUC     & 0.6888 & \textbf{0.7073} & 0.9175          & 0.9175          \\
  & Hits@1   & 0.3684 & \textbf{0.4211} & 0.4211          & 0.4211          \\
  & Hits@3   & 0.5263 & 0.5263          & 0.5263          & 0.5263          \\
  & Hits@5   & 0.5789 & \textbf{0.6316} & \textbf{0.6316} & 0.5789          \\
  & Hits@10  & 0.6842 & \textbf{0.7368} & 0.7368          & 0.7368          \\
  & MRR     & 0.4630 & \textbf{0.5124} & \textbf{0.5004} & 0.5002          \\
  & nDCG@10 & 0.4762 & \textbf{0.5331} & 0.5131          & \textbf{0.5149} \\
\midrule
\multirow{7}{*}{196}
  & AUC     & \textbf{0.7107} & 0.7054          & \textbf{0.9122} & 0.9119          \\
  & Hits@1   & \textbf{0.2105} & 0.1579          & 0.2368          & 0.2368          \\
  & Hits@3   & 0.3421          & \textbf{0.3684} & \textbf{0.3947} & 0.3684          \\
  & Hits@5   & 0.4474          & \textbf{0.4737} & 0.4211          & \textbf{0.4737} \\
  & Hits@10  & 0.6316          & \textbf{0.6579} & 0.6316          & \textbf{0.7105} \\
  & MRR     & \textbf{0.3215} & 0.3107          & \textbf{0.3548} & 0.3525          \\
  & nDCG@10 & 0.3737          & \textbf{0.3758} & 0.3922          & \textbf{0.4089} \\
\midrule
\multirow{7}{*}{351}
  & AUC     & 0.7258 & \textbf{0.7466} & 0.9246          & \textbf{0.9350} \\
  & Hits@1   & 0.0938 & \textbf{0.1094} & 0.0938          & 0.0938          \\
  & Hits@3   & 0.2344 & 0.2344          & 0.2656          & 0.2656          \\
  & Hits@5   & 0.3281 & 0.3281          & \textbf{0.3438} & 0.3281          \\
  & Hits@10  & 0.4531 & \textbf{0.5469} & 0.5156          & \textbf{0.5312} \\
  & MRR     & 0.2013 & \textbf{0.2194} & 0.2191          & \textbf{0.2193} \\
  & nDCG@10 & 0.2399 & \textbf{0.2764} & 0.2760          & \textbf{0.2765} \\
\bottomrule
\end{tabular}
\end{table}

The sensitivity analysis reveals three findings.

\paragraph{First, the biased random walk is sensitive to weight configuration, but the effect magnitude is modest.}
Across the three dataset sizes, genre emphasis shifts biased random walk ranking metrics by an average of approximately 3 percentage points, with the largest individual shift observed at Hits@10 on the 351-text configuration (a 9.4-point increase). The direction of the effect is not uniform: genre emphasis improves biased random walk on most metrics at 98 and 351 texts, but slightly degrades MRR and Hits@1 at 196 texts. This suggests that there is no single optimal weight configuration across dataset scales, and that weight selection should be treated as a tunable aspect of the recommendation pipeline rather than a fixed property of the domain.

\paragraph{Second, the hybrid model absorbs most of this sensitivity.}
Most hybrid metrics shift by less than 2 percentage points across weight configurations, with an average shift of approximately 1.5 percentage points across all ranking metrics and scales — roughly half the average shift observed for the biased random walk. The largest individual hybrid shifts (up to about 8 percentage points on Hits@10 at 196 texts) are smaller in magnitude than the corresponding biased random walk shifts and do not consistently favor either weight configuration. This dampening occurs because the hybrid embedding concatenates the biased random walk component with DeepWalk embeddings, which are unaffected by the weighting scheme. The DeepWalk component provides a stable structural backbone that reduces the biased random walk's sensitivity to weight perturbations, offering a practical advantage for deployment: hybrid performance is robust to reasonable variation in domain-expert weight choices.

\paragraph{Third, genre emphasis improves AUC on the larger graphs.}
Biased random walk AUC increases from 0.7258 to 0.7466 at 351 texts with genre emphasis, and hybrid AUC increases from 0.9246 to 0.9350. This effect, though small, is consistent with the interpretation that, on larger, sparser graphs, emphasizing a discrete categorical feature like genre helps the walk to concentrate on higher-signal neighborhoods, whereas on smaller, denser graphs, the default balanced weighting already captures the same information through the overall topology.

\subsection{Qualitative Case Analysis}
\label{sec:case-analysis}

To complement the quantitative metrics with a concrete illustration of system behavior, Table~\ref{tab:case-1984} presents the top-10 recommendations generated by each of the six embedding configurations for the anchor text \textit{1984} by George Orwell, drawn from the 351-text dataset run. The five expert-curated ground-truth recommendations for this anchor are \textit{Fahrenheit 451}, \textit{Brave New World}, \textit{Animal Farm}, \textit{The Hunger Games}, and \textit{Marrow Thieves}, connected through themes of dystopian fiction, censorship, oppression, totalitarianism, and rebellion. \textit{1984} was selected for this case study because it had the highest average ground-truth recall across all six configurations among the 64 anchor texts in the evaluation set, and because its recommendations exhibit clear inter-model variation that illustrates how different embedding strategies surface different but pedagogically defensible associations.

The recommendations are presented as two halves of a single table. The upper half compares DeepWalk, Node2Vec, and biased random walk under default weights; the lower half compares biased random walk under genre-emphasized weights with both hybrid configurations. Bolded entries indicate matches with the expert-curated ground-truth set.

\begin{table}[htbp]
\centering
\caption{Top-10 recommendations from \mbox{T-TExTS} for the anchor text \textit{1984} across all six embedding configurations on the 351-text dataset. Bolded entries match the expert-curated ground-truth set: \textit{Fahrenheit 451}, \textit{Brave New World}, \textit{Animal Farm}, \textit{The Hunger Games}, \textit{Marrow Thieves}.}
\label{tab:case-1984}
\small
\setlength{\tabcolsep}{5pt}
\renewcommand{\arraystretch}{1.25}
\begin{tabular}{c >{\raggedright\arraybackslash}p{3.1cm} >{\raggedright\arraybackslash}p{3.1cm} >{\raggedright\arraybackslash}p{3.1cm}}
\toprule
\multicolumn{4}{c}{\textbf{(a) Default-weight configurations}} \\
\midrule
\textbf{Rank} & \textbf{DeepWalk} & \textbf{Node2Vec} & \textbf{Biased RW (default)} \\
\midrule
1  & \textbf{\textit{Fahrenheit 451}}    & \textbf{\textit{Fahrenheit 451}}     & \textbf{\textit{Fahrenheit 451}}     \\
2  & \textbf{\textit{Brave New World}}   & \textbf{\textit{Brave New World}}    & \textit{The Giver}                   \\
3  & \textbf{\textit{Animal Farm}}       & \textit{Scythe}                      & \textbf{\textit{Brave New World}}    \\
4  & \textit{My Name is Asher Lev}       & \textit{The Giver}                   & \textit{And Then There Were None}    \\
5  & \textit{And Then There Were None}   & \textbf{\textit{The Hunger Games}}   & \textit{The Pedestrian}              \\
6  & \textbf{\textit{Marrow Thieves}}    & \textbf{\textit{Marrow Thieves}}     & \textit{The Road}                    \\
7  & \textit{The Giver}                  & \textit{A Bot Might Have Written This} & \textit{Scythe}                    \\
8  & \textit{The Last Dog}               & \textit{The Pedestrian}              & \textit{The Martian}                 \\
9  & \textit{The Star}                   & \textit{The Martian}                 & \textit{The Last Dog}                \\
10 & \textit{The Pedestrian}             & \textbf{\textit{Animal Farm}}        & \textbf{\textit{The Hunger Games}}   \\
\midrule
\textbf{Hits / 5} & 4 & 5 & 3 \\
\midrule
\midrule
\multicolumn{4}{c}{\textbf{(b) Genre-emphasized and hybrid configurations}} \\
\midrule
\textbf{Rank} & \textbf{Biased RW (genre-emp.)} & \textbf{Hybrid (default)} & \textbf{Hybrid (genre-emp.)} \\
\midrule
1  & \textbf{\textit{Fahrenheit 451}}              & \textbf{\textit{Fahrenheit 451}}    & \textbf{\textit{Fahrenheit 451}}    \\
2  & \textit{The Giver}                            & \textbf{\textit{Brave New World}}   & \textbf{\textit{Brave New World}}   \\
3  & \textit{The Pedestrian}                       & \textit{The Giver}                  & \textit{And Then There Were None}   \\
4  & \textit{And Then There Were None}             & \textit{And Then There Were None}   & \textit{The Giver}                  \\
5  & \textbf{\textit{Brave New World}}             & \textbf{\textit{Animal Farm}}       & \textit{The Pedestrian}             \\
6  & \textit{The Last Dog}                         & \textit{The Pedestrian}             & \textbf{\textit{Animal Farm}}       \\
7  & \textit{The Martian}                          & \textit{Scythe}                     & \textit{The Last Dog}               \\
8  & \textit{Scythe}                               & \textit{The Last Dog}               & \textit{The Martian}                \\
9  & \textbf{\textit{The Hunger Games}}            & \textit{The Martian}                & \textit{Scythe}                     \\
10 & \textit{The Immortal Life of Henrietta Lacks} & \textbf{\textit{The Hunger Games}}  & \textbf{\textit{The Hunger Games}}  \\
\midrule
\textbf{Hits / 5} & 3 & 4 & 4 \\
\bottomrule
\end{tabular}
\end{table}

Four patterns emerge from these recommendations.

\paragraph{All six configurations rank \textit{Fahrenheit 451} first.}
This unanimous top placement reflects the strong structural alignment between \textit{1984} and \textit{Fahrenheit 451} in the knowledge graph: the two texts share genre (dystopian fiction), multiple themes (censorship, oppression, totalitarianism), and comparable text-complexity ratings. The consistency across models, despite their differing traversal strategies, indicates that the strongest pedagogical signals are robustly captured by the underlying ontology and surface regardless of the embedding method chosen.

\paragraph{Node2Vec is the only configuration to surface all five ground-truth texts.}
Node2Vec recovers \textit{Fahrenheit 451}, \textit{Brave New World}, \textit{The Hunger Games}, \textit{Marrow Thieves}, and \textit{Animal Farm} within its top 10. This reflects the same pattern observed quantitatively in Tables~\ref{tab:main-results} and \ref{tab:sensitivity}, where Node2Vec's parameterized traversal (with the consistently selected $q=4.0$, encouraging DFS-biased exploration) yields the strongest ranking metrics at the 351-text scale. By moving more aggressively away from the immediate neighborhood of the anchor, Node2Vec surfaces structurally distant but thematically related texts that the more locally-bounded traversal of biased random walks tends to overlook.

\paragraph{The hybrid model recovers ground-truth texts that a biased random walk alone misses.}
While the biased random walk (default) recovers only three ground-truth texts (\textit{Fahrenheit 451}, \textit{Brave New World}, \textit{The Hunger Games}), the hybrid configuration adds \textit{Animal Farm} to the recovered set. This gain comes from the DeepWalk component of the hybrid embedding, which contributes broader topological coverage and surfaces \textit{Animal Farm}, an Orwell-authored text that biased random walk's narrower exploration tends to push outside the top 10. The hybrid model thus inherits structural breadth from DeepWalk and pedagogical bias from the biased random walk component, consistent with the AUC-versus-ranking trade-off behavior reported quantitatively.

\paragraph{Strong near-miss recommendations are consistent across configurations.}
Several texts that are not in the ground-truth set appear in nearly every model's top 10, including \textit{The Giver}, \textit{The Pedestrian}, and \textit{Scythe}. These are dystopian narratives in young-adult and short-fiction formats that share the anchor's central concerns of conformity, surveillance, and individual autonomy. From a pedagogical standpoint, such near-miss recommendations are not failures of the model, but rather, legitimate alternatives that a teacher might consider when assembling a text set, particularly one designed to introduce dystopian themes through varied lengths and reading levels. This reinforces the broader observation from Section~\ref{sec:results-main} that ranking metrics, which penalize any deviation from the expert-curated ground truth, may underrepresent the practical utility of the system for curriculum design.

Taken together, the case analysis illustrates how \mbox{T-TExTS} produces recommendations that are simultaneously \textit{diverse} (spanning multiple formats and reading levels within the broader dystopian and oppression-themed space) and \textit{pedagogically aligned} (consistently surfacing texts that share genre, theme, and thematic concerns with the anchor). This dual property, which is the central design goal of the system, is robust across all six embedding configurations evaluated.

\section{Discussion and Limitations}
\label{sec:discussion_and_limitation}

\subsection{Discussion}
\label{sec:discussion}
The central empirical finding of this study is that traversal-level expert weighting alone does not outperform algorithmic structural tuning. Across the three dataset sizes (98, 196, and 351 high school English Literature texts), the biased random walk achieves neither the highest AUC (0.6888--0.7466) nor the strongest results on any ranking metric. Node2Vec, in contrast, achieves the highest AUC at every scale (0.9642--0.9750) and the strongest ranking metrics on the larger graphs. The hybrid embedding, which concatenates DeepWalk and biased random walk representations, retains a high AUC across all scales (0.9122--0.9350) while remaining within a few percentage points of Node2Vec on every ranking metric, and additionally reveals which expert-assigned weights influenced the result. For a teacher-facing tool where transparency matters, this combination of competitive performance and traceable pedagogical input makes the hybrid model the most practical deployment choice.

To interpret this result, it is useful to examine the role of pedagogy in the \mbox{T-TExTS} recommendation pipeline. In \mbox{T-TExTS}, expert input shapes the recommendation pipeline at two distinct layers. The first layer is the knowledge graph itself: the ontology schema and the per-text annotations were both curated by domain experts following the KNARM methodology, so every relation, entity, and attribute that any model traverses is already pedagogically structured. The second layer is the traversal step: the biased random walk further amplifies expert-assigned weights on specific relations such as \texttt{has\_genre}, \texttt{has\_theme}, and \texttt{has\_subtheme}, while DeepWalk and Node2Vec operate on the same graph without applying additional weighting at traversal time. All four models, therefore, inherit the graph-level pedagogy; only the biased random walk and the hybrid model add the traversal-level weighting. The empirical pattern indicates that the pedagogical signal is already captured strongly enough by the graph structure that additional traversal-level weighting is not strictly necessary, and may even reduce performance by sacrificing the structural coverage that AUC rewards. This validates the ontology design: the graph carries the pedagogical signal strongly enough that a structurally-tuned model can recover it without further intervention.

Two further observations support this finding. First, AUC is remarkably stable across dataset sizes: DeepWalk varies only between 0.9595 and 0.9653, and Node2Vec between 0.9642 and 0.9750, even as the graph triples in entity count from 868 to 1,312 unique entities. The expert-curated structure, therefore, generalizes from smaller to larger graphs, an important property for any ontology-driven recommendation system intended to scale with curriculum expansion. Second, no single embedding strategy dominates every metric at every scale. DeepWalk leads ranking at the 98-text scale, where the graph is small and densely connected, and uniform exploration captures the relevant pedagogical neighborhoods directly. As the graph grows and candidates compete more sharply, Node2Vec's parameterized traversal takes over on the ranking side. The hybrid model is consistently second-best on most ranking metrics, which makes it the best practical choice when both AUC and ranking quality matter.

A specific pattern in the Node2Vec hyperparameters warrants closer examination. Across all three dataset sizes, Optuna independently selected $q=4.0$, the maximum value in our search space, while $p$ varied between 0.25 and 2.0. A high $q$ in Node2Vec discourages the walk from returning to neighborhoods it has already visited, biasing exploration toward depth-first traversal that moves outward from the source node. That this preference holds at every scale is itself informative: it suggests that the most useful pedagogical signal in our graph is found in longer chains of association rather than in tightly clustered local neighborhoods.

This pattern reflects a property of the graph design itself. The KNARM ontology deliberately connects texts through intermediate concepts such as Genre, Theme, Subtheme, and qualitative measures, so a walk that pushes outward through these intermediates surfaces texts that share the anchor's deeper pedagogical concerns through indirect routes. A walk that lingers near its starting point will overweight surface similarity (e.g., other works by the same author) at the expense of pedagogically aligned but structurally distant texts. The case-study finding supports this interpretation: Node2Vec was the only configuration to recover all five ground-truth texts for \textit{1984}, including \textit{Marrow Thieves}, which is structurally distant but thematically aligned through dystopian rebellion narratives. The implication for future work on KG-based literature recommendation is that traversal strategies that actively encourage broad thematic exploration may yield more pedagogically relevant recommendations than strategies that preserve local clustering, particularly when the graph is built for pedagogical coverage rather than user-similarity signals.

The sensitivity analysis in Section~\ref{sec:sensitivity} reinforces a deployment-relevant point. The biased random walk shifts by an average of approximately 3 percentage points on ranking metrics when the genre weight is increased from 3 to 4, with the largest individual shift reaching 9.4 points (Hits@10 at 351 texts). The direction of the shift is not consistent across dataset sizes. The hybrid model exhibits roughly half of this average sensitivity (about 1.5 percentage points across ranking metrics), and most of its individual metric shifts remain under 2 percentage points. This dampening is attributable to the hybrid's DeepWalk component, which is weight-independent and contributes a stable structural backbone that reduces the biased random walk's sensitivity to weight perturbations. In a setting where domain-expert weights may legitimately differ between teaching contexts, this robustness is a practical advantage: a hybrid-based deployment will produce reasonably similar recommendations even when the underlying weight scheme is adjusted.

\subsection{Limitations}
\label{sec:limitations}

Despite the strong quantitative results obtained by \mbox{T-TExTS} on the secondary-level English Literature text recommendation task, several limitations should temper the interpretation of these findings.

Although the dataset of 351 texts represents a substantial body of expert-curated content for the secondary-level English Literature domain, it remains modest in scale relative to general-purpose recommendation benchmarks. The scaling study reported in Section~\ref{sec:results-main} indicates that the proposed embedding methods generalize stably as the graph grows from 98 to 351 texts, but extrapolation beyond this range is not directly evidenced by the present experiments.

We did not conduct a live user study with classroom teachers in this work. The evaluation throughout this paper relies on expert-curated ground-truth recommendations rather than measurements of how the system performs in actual curriculum-planning workflows. Teacher-facing validation, including measurements of planning-time savings and text-set diversity in real classrooms, remains future work.

The ground-truth set itself reflects the judgment of a single expert panel and not a consensus across multiple panels. A different group of educators might identify a different set of pedagogically-aligned texts for the same anchor, particularly for cases where the anchor admits multiple defensible thematic readings. The case study makes this concrete: several of the strongest near-miss recommendations (\textit{The Giver}, \textit{The Pedestrian}, \textit{Scythe}) are pedagogically plausible companions for \textit{1984} but were not in our ground-truth set, so models that surface them are penalized by Hits@K even though a teacher might rank them highly.
 
This points to a deeper limitation of the evaluation methodology. Standard ranking metrics (Hits@K, MRR, nDCG) were developed for large-scale, sparse recommendation graphs in which the user's preferred items are well-defined and reasonably stable. They do not naturally accommodate the kind of pedagogical near-misses that surface in expert-curated, neuro-symbolic settings, nor do they reward recommendations that are structurally distant but thematically apt. New evaluation metrics for expert-driven knowledge graphs are needed, ideally ones that incorporate ontological proximity, pedagogical equivalence classes, and weighted partial credit for near-misses, rather than treating recommendation as a binary match against a fixed ground-truth list. We see this as a broader open problem for neuro-symbolic recommendation research and not specific to the present work on \mbox{T-TExTS}.

Finally, the biased random walk weight scheme was evaluated under two configurations (default and genre-emphasized) rather than searched exhaustively. A more thorough exploration of the weight space, possibly using Bayesian optimization over the weight assignments themselves, may identify configurations that further improve biased-walk performance. We treat the two configurations reported here as a sensitivity check, not as a claim that the weights are optimal.

\section{Conclusion and Future Work}
\label{sec:conclusion_and_future_work}

\subsection{Conclusion}
\label{sec:conclusion}

This paper introduced \mbox{T-TExTS}, a recommendation system that assists high school English Literature teachers in selecting texts that are diverse in genre, theme, subtheme, and author, yet similar in pedagogical merits, using a domain-specific knowledge graph. The system makes three contributions, each of which we treat as essential to the overall claim. First, a pedagogy-grounded ontology was constructed using the KNARM methodology, capturing not only genres and themes but also instructional and qualitative pedagogical elements (levels of meaning, text structure, language conventionality and clarity, knowledge demands) that prior text-recommendation systems have not modeled together. Second, this ontology was instantiated as a knowledge graph and used to drive a recommendation pipeline based on random-walk graph embeddings. Third, we conducted a comparative evaluation of four embedding strategies (DeepWalk, biased random walk, hybrid, and Node2Vec) across three dataset sizes (98, 196, and 351 texts) and two weight configurations, yielding the main empirical findings of this paper.

The central empirical finding is that traversal-level expert weighting alone does not outperform algorithmic structural tuning on either AUC or any ranking metric, but combining the two preserves both pedagogical interpretability and competitive ranking quality. Node2Vec, which operates on the same expert-curated graph as the biased random walk but applies no additional traversal-level weighting, achieves the highest AUC at every dataset size (0.9642 to 0.9750) and the strongest ranking metrics on the larger graphs. The hybrid embedding, which concatenates DeepWalk and biased random walk representations, retains a high AUC across all scales (0.9122 to 0.9350) while staying within a few percentage points of Node2Vec on every ranking metric. This makes the hybrid model the most practical choice for deployment in a teacher-facing tool: it inherits the structural robustness of uniform exploration while exposing which expert-assigned weights influenced the result, providing the transparency that classroom-facing applications typically require.

These findings extend prior work on KG-based recommendation in a specific direction. They demonstrate that high-fidelity, expert-curated knowledge graphs offer a viable foundation for specialized educational scaffolding tools at the modest data scales typical of curriculum-domain ontologies, and that the relative ordering of embedding methods that we observed is stable across three dataset sizes ranging from 98 to 351 texts. \mbox{T-TExTS} therefore provides both a concrete tool for English Literature text selection and a methodological template for building pedagogy-aware recommendation systems in other curriculum domains where expert input is more abundant than user-interaction data.

\subsection{Future Work}
\label{sec:future_work}

Several directions follow naturally from this work.

The most immediate is teacher-facing validation. While the present evaluation relies on expert-curated ground-truth recommendations, the practical value of \mbox{T-TExTS} depends on whether the system reduces planning time and improves text-set diversity in real-world classrooms. We plan to conduct a user study with secondary English educators, measuring how the recommendations are received in curriculum-design workflows and whether the diversity-yet-coherence property of \mbox{T-TExTS} surfaces as predicted.

A second direction concerns evaluation methodology. The Limitations section highlighted that standard ranking metrics (Hits@K, MRR, nDCG) penalize pedagogically defensible near-misses and treat recommendations as a binary match against a fixed ground-truth list. However, for expert-driven, neuro-symbolic recommendation systems, new metrics are needed that incorporate ontological proximity, pedagogical equivalence classes, and weighted partial credit for structurally distant but thematically apt recommendations. We view this as a broader research direction relevant to neuro-symbolic AI as a whole, and we plan to operationalize and test such metrics on \mbox{T-TExTS} as a starting case.

Finally, integrating \mbox{T-TExTS} with retrieval-augmented generation (RAG) presents a promising extension. GraphRAG \cite{graphrag}, which combines knowledge graphs with large language models (LLMs), would allow teachers to ask natural-language questions about texts and receive grounded, ontology-informed answers. This would extend the \mbox{T-TExTS} system's role from recommender to interactive curriculum-planning assistant, while preserving the pedagogical grounding that the underlying ontology provides.

\backmatter

\bibliography{sn-bibliography}

\end{document}